\documentclass[a4paper,11pt]{article}
\usepackage{pos}
\usepackage{physics}
\usepackage{tensor}
\usepackage{subcaption}
\usepackage{graphicx}
\usepackage{placeins}
\title{Shaken, not stirred: Test particles in binary black hole mergers}

\author*[a]{Pieter vd Merwe}
\author[a]{Markus B\"ottcher}

\affiliation[a]{Centre for Space Research, North-West University,\\
                Potchefstroom, 2520, South Africa\\}

\emailAdd{pietervdmerwejnr@gmail.com}
\emailAdd{Markus.Bottcher@nwu.ac.za}

\abstract{Since 2015 the advanced Laser Interferometer Gravitational-Wave Observatory (aLIGO) has detected a large number of gravitational wave events, originating from both binary neutron stars and binary black hole (BBH) mergers. In light of these detections, we simulate the dynamics of ambient test particles in the gravitational potential well of a BBH system close to its inspiral phase with the goal of simulating the associated electromagnetic radiation and resulting spectral energy distribution of such a BBH system. This could shed light on possible detection ranges of electromagnetic counterparts to BBH mergers. The potentials are numerically calculated using finite difference methods, under the assumption of non-rotating black holes with the post-Newtonian Paczynski-Wiita potential approximation in tandem with retarded time concepts analogous to electrodynamics. We find that the frequencies of potential electromagnetic radiation produced by these systems (possibly reaching Earth), range between a few $\text{kHz}$ to a few $100 \text{MHz}$.}

\FullConference{%
  
High Energy Astrophysics in Southern Africa 2022 - HEASA2022\\
  28 September - 1 October 2022\\
  Brandfort, South Africa
}

\begin{document}
\raggedbottom
\maketitle

\section{Introduction} \label{sec:intro}
Since the detection of the first gravitational wave (GW) event in 2015 by aLIGO \citep{GWFirst1,GWFirst2}, multiple binary black hole (BBH), binary neutron star (BNS), and black hole - neutron star mergers have been detected \citep{aLIGOO31,LIGOO32,Abbott_2019,PhysRevX.9.031040,Abbott_2019_2,Abbott_2021}, with many new prospects for future detectors such as the laser interferometer space antenna (LISA) \citep{larson2005lisa, bassan2017lisa} and the Einstein telescope \citep{Maggiore_2020,Punturo_2010}.  With these detections, a new window on multi-messenger astronomy has opened and the search for multiple messengers (gravitational waves, particles, and photons) from various sources began \citep{Abbott_2017,Abbott_2019_3,Abbott_2022_2,PhysRevD.89.122004,GWGRB1,Valenti_2017,FRB1,Abbott_2021_SN,Aasi_2014}.

Coincident with the detection of the first GW event, the \textit{Fermi} gamma-ray burst monitor (GBM) detected a weak transient gamma-ray burst (GRB) signal (GW150914-GBM) of 1s duration, 0.4s after the BBH merger event GW150914, with a consistent sky localization \citep{Connaughton_2016, Connaughton_2018}. The GRB was found during an off-line analysis of the Fermi GBM data \citep{GBM1, Kocevski_2018}. This possible electromagnetic (EM) counterpart to a BBH merger event is highly unexpected and, as such, led to many speculations and counter arguments regarding possible scenarios that could produce observable EM signals from a BBH merger \citep{Loeb_2016,Dai_2017,Fedrow_2017,Perna_2016,Woosley_2016,Kimura_2016,Zhang_2016}, and the possible GRB applicaption thereof should the detection be valid \citep{Veres_2016,Veres_2019}. It is believed that GW events resulting from stellar-mass BBH mergers are unlikely to have EM counterparts \citep{de2017electromagnetic}. However, the detection of GW150914-GBM encouraged theoretical speculation on possible scenarios in which EM counterparts to GW signals are possible.

In light of these findings, we develop a na\"ive, pseudo-Newtonian exploratory code using the Paczynski-Wiita potential and retarded time concept to investigate accelerations and possible spectral energy distribution (SED) forms obtainable from evolving ambient particles present in a BBH merger system (in the inspiral phase) under the influence of gravity. Since this is intended as a na\"ive exploratory code to possibly be further built upon later, the addition of accretion disk considerations and fully general relativistic magnetohydrodynamic simulations such as those done by \citep{Khan2018} have been neglected in this work. In section \ref{chpt:Mod}, we give a brief overview of the building blocks used to develop the model and the code. Section \ref{chpt:Results} shows a set of test results obtained from the code. A discussion on the test results and future prospects is given in section  \ref{chpt:Discussion}.

\section{\label{chpt:Mod}The model}

This section describes the model used to determine the gravitational acceleration of a charged particle in an inspiral-phase BBH system. The gravitational acceleration of the particle is determined using the Paczynski-Wiita gravitational potential, as described in Section \ref{Subsec:PacWii}. Additionally, we introduce the concept of retarded time (analogous to electrodynamics) in order to determine the acceleration of the particle.    

The effective one-body (EOB) problem entails treating two bodies, with masses $M_{1}$ and $M_{2}$, with a separation $\vec{r}=\vec{r}_{2}-\vec{r}_{1}$, orbiting a common center of mass as a single body with a reduced mass $\mu=\frac{M_{1}M_{2}}{M_{1}+M_{2}}$ under the influence of an external potential due to a total mass $M=M_{1}+M_{2}$. To first post-Newtonian order, the analytical solution of the circularized, inspiral BBH orbit can be derived as
\begin{minipage}[b]{0.4\columnwidth}
\begin{equation}\label{eq:BBHPosAsFuncTime}
\resizebox{0.7\textwidth}{!}{$R\left(t\right) = R_{0}\left(\frac{t_{c}-t}{t_{c}}\right)^{-0.25}\frac{M_{c}}{M},$}
\end{equation}  
\end{minipage}
\begin{minipage}[b]{0.5\columnwidth}
\begin{equation}\label{eq:BBHPosAsFuncTime2}
\resizebox{0.85\textwidth}{!}{$\text{and, \hspace{12pt}}\Phi\left(t\right) = -2\left(5M_{c}\right)^{-0.625}\left(t_{c}-t\right)^{-0.25},$}
\end{equation}
\end{minipage}

\noindent with $t_{c}$ the time to coalescence of the BHs, $R_{0}$ the initial separation of the BHs, $M_{c}\equiv\frac{\left(M_{1}M_{2}\right)^{0.6}}{\left(M_{1}+M_{2}\right)^{0.2}}$ the chirp mass of the BBHs, $R\left(t\right)$ the separation as a function of time and $\Phi\left(t\right)$ the dimensionless phase of $\mu$ in the center of momentum frame \citep{maggiore2007gravitational}. From this the position vectors $\vec{r}_{M_{1}}$ and $\vec{r}_{M_{2}}$ can be found similarly to the classical EOB problem.

\subsection{\label{Subsec:PacWii}The Paczynski-Wiita Potential} 
The Paczynski-Wiita potential is an important pseudo-Newtonian approximation to the general relativistic Schwarzschild geometry developed by B. Paczynski and P. Wiita while studying thick accretion disks and super-Eddington luminosities \citep{PacWiiOG}:
\begin{equation}\label{eq:PacWii1}
\begin{split}
\Phi_{PW}\left(r\right) &= -\frac{GM}{r-r_{s}},\ \ \ \ r_{s}=\frac{2GM}{c^{2}}.        \\
\end{split}
\end{equation}
The Paczynski-Wiita potential is a Newton-like potential that (for the case of non-rotating BHs) exactly reproduces the marginally bound ($r_{mb}=2r_{s}$) and marginally stable circular ($r_{ms}=3r_{s}$) orbits of the Schwarzschild geometry, as well as the form of the Keplerian angular momentum, but not that of the Keplerian angular velocity \citep{PacWii,torkelsson1998special}. \citep{Witzany2015} has shown that orbits resulting from a Paczynski-Wiita potential exhibit satisfactory correspondence with the exact General relativistic (GR) case, with the caveat that these orbits are likely to be more chaotic than those expected from a fully relativistic approach. Hence, any predictions made in this article regarding particle acceleration and electromagnetic radiation may be considered as upper limits to the radiative output in full GR.

\subsection{\label{sec:particleCalculations}Calculating the acceleration of a charged particle in an inspiral phase BBH merger}
Considering a charged particle, with mass $m_{e}$, position $\vec{r}$, and velocity $\vec{v}$ in the lab frame, if the particle is under the influence of a force $\vec{F}=\dv{\vec{p}}{t}$, then to determine $\vec{F}$, the retarded positions of the BHs, with respect to the position of the charged particle, have to be determined first. For this purpose, we define retarded time $t_{r}\equiv t-\frac{1}{c}\left|\vec{r}-\vec{r}_{k}\right|$, where $\vec{r}$ is the position of the particle and $\vec{r}_{k}$ is the position of the BH at time $t_{r}$. The retarded times, $t_{r,M_{1}}$ and $t_{r,M_{2}}$, at which to evaluate the position vectors of the two BHs are determined using the Newton-Rhapson numerical root-finding method.

Once we have the retarded times $t_{r,M_{1}}$ and $t_{r,M_{2}}$, $\vec{F}$ can be determined. Using the Paczynski-Wiita potential, as described in Section \ref{Subsec:PacWii}, where $\Phi_{PW,1}$ and $\Phi_{PW,2}$ denotes the potential resulting from $M_{1}$ and $M_{2}$ respectfully, we have that 
\begin{equation}\label{eq:FiPacWii}
\begin{split}
\vec{F} &= m_{e}\vec{\nabla}\left[\Phi_{PW,1}\left(\left|\vec{r}-\vec{r}'_{1}\left(t_{r,M_{1}}\right)\right|\right)+\Phi_{PW,2}\left(\left|\vec{r}-\vec{r}'_{2}\left(t_{r,M_{2}}\right)\right|\right)\right].        \\
\end{split}
\end{equation}
Furthermore, it follows that Equation \ref{eq:FiPacWii} simplifies to
\begin{equation}\label{eq:FinalFiPacWii}
\resizebox{0.9\hsize}{!}{$\vec{F} = -m_{e}\left[\frac{GM_{1}}{r_{d,1}\left(r_{d,1}-r_{s,1}\right)^{2}}\left(x_{d,1}\hat{x}+y_{d,1}\hat{y}+z_{d,1}\hat{z}\right)+\frac{GM_{2}}{r_{d,2}\left(r_{d,2}-r_{s,2}\right)^{2}}\left(x_{d,2}\hat{x}+y_{d,2}\hat{y}+z_{d,2}\hat{z}\right)\right],$}
\end{equation}
where \resizebox{0.19\hsize}{!}{$r_{d,k}=\left|\vec{r}-\vec{r}'_{k}\left(t_{r,M_{k}}\right)\right|$}, \resizebox{0.19\hsize}{!}{$x_{d,k}=\left(\vec{r}-\vec{r}'_{k}\left(t_{r,M_{k}}\right)\right)_{x}$}, 
\resizebox{0.19\hsize}{!}{$y_{d,k}=\left(\vec{r}-\vec{r}'_{k}\left(t_{r,M_{k}}\right)\right)_{y}$},  \resizebox{0.19\hsize}{!}{$z_{d,k}=\left(\vec{r}-\vec{r}'_{k}\left(t_{r,M_{k}}\right)\right)_{z}$}, and $k=1,2$.
From the definition $\vec{p}=\gamma m_{e}\vec{v}$ we have that
\begin{equation}\label{eq:DetVi2}
 v^{2} = \frac{p^{2}}{m_{e}^{2}}\left(1+\frac{p^{2}}{m_{e}^{2}c^{2}}\right)^{-1},\; \; \text{\vspace{4pt}and\vspace{12pt}}\; \; \vec{v} = \frac{\vec{p}}{m_{e}}\sqrt{1-\frac{p^{2}}{m_{e}^{2}c^{2}}\left(1+\frac{p^{2}}{m_{e}^{2}c^{2}}\right)^{-1}}.
\end{equation}

For a given velocity we can calculate the position from the definition of velocity, $\vec{v} = \dv{\vec{r}}{t}.$ Substituting equation \ref{eq:DetVi2} into this and converting to geometrized units\footnote{We convert to geometrized units here for better numerical accuracy as terms containing $G$ and $c$ can suffer from numerical rounding errors.} we find a set of coupled first-order ordinary differential equations (ODEs),
\begin{equation}\label{eq:ODEs}
\dv{\vec{p}}{t} = \vec{F}\left(t,\vec{r}\right),    \text{\hspace{16pt} and, \hspace{16pt}}
\dv{\vec{r}}{t} = \frac{\vec{p}}{m_{e}}\sqrt{1-\frac{p^{2}}{m_{e}^{2}}\left(1+\frac{p^{2}}{m_{e}^{2}}\right)^{-1}},\\
\end{equation}
that must be solved simultaneously, with $\vec{F}$ calculated as described in equation \ref{eq:FinalFiPacWii}. In order to simultaneously solve this set of ODEs we use the embedded 7(8) Runge Kutta method devised by \cite{prince1981high}.
In order to calculate the acceleration $\vec{a}$ of the particle under the influence of the retarded force $\vec{F}$, the equation 
\begin{equation}\label{eq:Relativistic3AccPart2}
\begin{split}
 \vec{a} &= \frac{1}{m\gamma}\left[\vec{F}-\left(\vec{v}\cdot\vec{F}\right)\vec{v}\right],\\
\end{split}
\end{equation}
is used.
Finally, the acceleration of the particle $\vec{a}$ is separated into its parallel, $\vec{a}_{\parallel} = \left(\vec{a}\cdot\frac{\vec{v}}{\left|\vec{v}\right|}\right)\hat{v}$, and perpendicular, $\vec{a}_{\perp} = \vec{a}-\vec{a}_{\parallel}$, components, with respect to the motion of the particle. 

\subsection{\label{sec:calculatingSEDs}Calculating the SED of charged particles in a near-inspiral BBH merger}
\noindent The SED of a single accelerated charged particle is given by
\begin{equation}\label{eq:SEDCh4}
\begin{split}
I\left(\nu\right) &= \frac{\mu_{0}q^{2}}{3\pi c}\left[\left|\gamma^{2}\left(\nu\right)\vec{a}_{\|}\left(\nu\right)\right|^{2}+\left|\gamma^{3}\left(\nu\right)\vec{a}_{\bot}\left(\nu\right)\right|^{2}\right].     \\
\end{split}
\end{equation}
From this, it is evident that $I_{\parallel,b}=\left|\gamma^{2}\left(\nu_{b}\right)\vec{a}_{\|}\left(\nu_{b}\right)\right|^{2}$, and $I_{\perp,b}=\left|\gamma^{3}\left(nu_{b}\right)\vec{a}_{\bot}\left(\nu_{b}\right)\right|^{2}$ needs to be determined in frequency space\footnote{Here the subscript $b$ only serves to indicate discretization.}. This is done by simply taking the discrete Fourier transform of the relevant products:\\
\begin{minipage}[b]{0.5\columnwidth}
\begin{equation}\label{eq:parallelDFT}
\begin{split}
I_{\parallel,b}&=\frac{1}{N}\sum_{j=0}^{N-1}\left|\gamma^{2}_{j}\vec{a}_{\|,j}\right|^{2} e^{-\frac{i2\pi}{N}bj},          \\
\end{split}
\end{equation}
\end{minipage}
\begin{minipage}[b]{0.5\columnwidth}
\begin{equation}\label{eq:perpDFT}
\begin{split}
I_{\perp,b}&=\frac{1}{N}\sum_{j=0}^{N-1}\left|\gamma^{3}_{j}\vec{a}_{\perp,j}\right|^{2} e^{-\frac{i2\pi}{N}bj}         , \\
\end{split}
\end{equation}
\end{minipage}
 with $i=\sqrt{-1}$, $j=0,1,..,N$ where $N$ is the number of uniformly distributed time domain data points, $b=0,1,...,N$ \citep{finufft1,finufft2}. In order to perform the fast Fourier transform (FFT), which requires a set of uniformly spaced data points, the non-uniformly spaced acceleration data is interpolated using cubic spline interpolation.
 
Equation \ref{eq:SEDCh4} gives the SED of a single particle. Considering a set of $N_{particles}$ particles, each with an SED $I_{n}\left(\nu\right)$, the total SED of the system can be found by summing over the SEDs of all of the particles in the system\footnote{Here $n = 1,..,N_{particles}$}.

Since it is expected that the radiation contribution of each particle is incoherent, the total SED can be calculated in this manner without the need to consider possible interference when summing over the fluxes of the particles.
When considering a set of discrete SEDs $I_{n}\left(\nu_{b}\right)$ for each particle, however, this process becomes somewhat more complicated. Applying the FFT in the calculation of the SEDs of each of the $N_{particles}$ particles, the implicit discrete frequency steps $d\nu$ within the FFTs are not necessarily of the same size.\footnote{This comes from the fact that the size $N$ of the input data set of the FFT will not be the same size for each particle and correspondingly $\nu_{b}$ (the set of discretized frequencies for each particle) will also be at different frequencies for each particle.} To account for this, each individual particle SED $I_{n}\left(\nu\right)$ is interpolated to comparable frequency increments, using cubic spline interpolation.


\section{\label{chpt:Results}Results} 

\begin{figure}[!hbt]%
    \centering%
    \begin{subfigure}{0.45\columnwidth}
        \resizebox{0.8\hsize}{!}{\includegraphics[scale = 1.0]{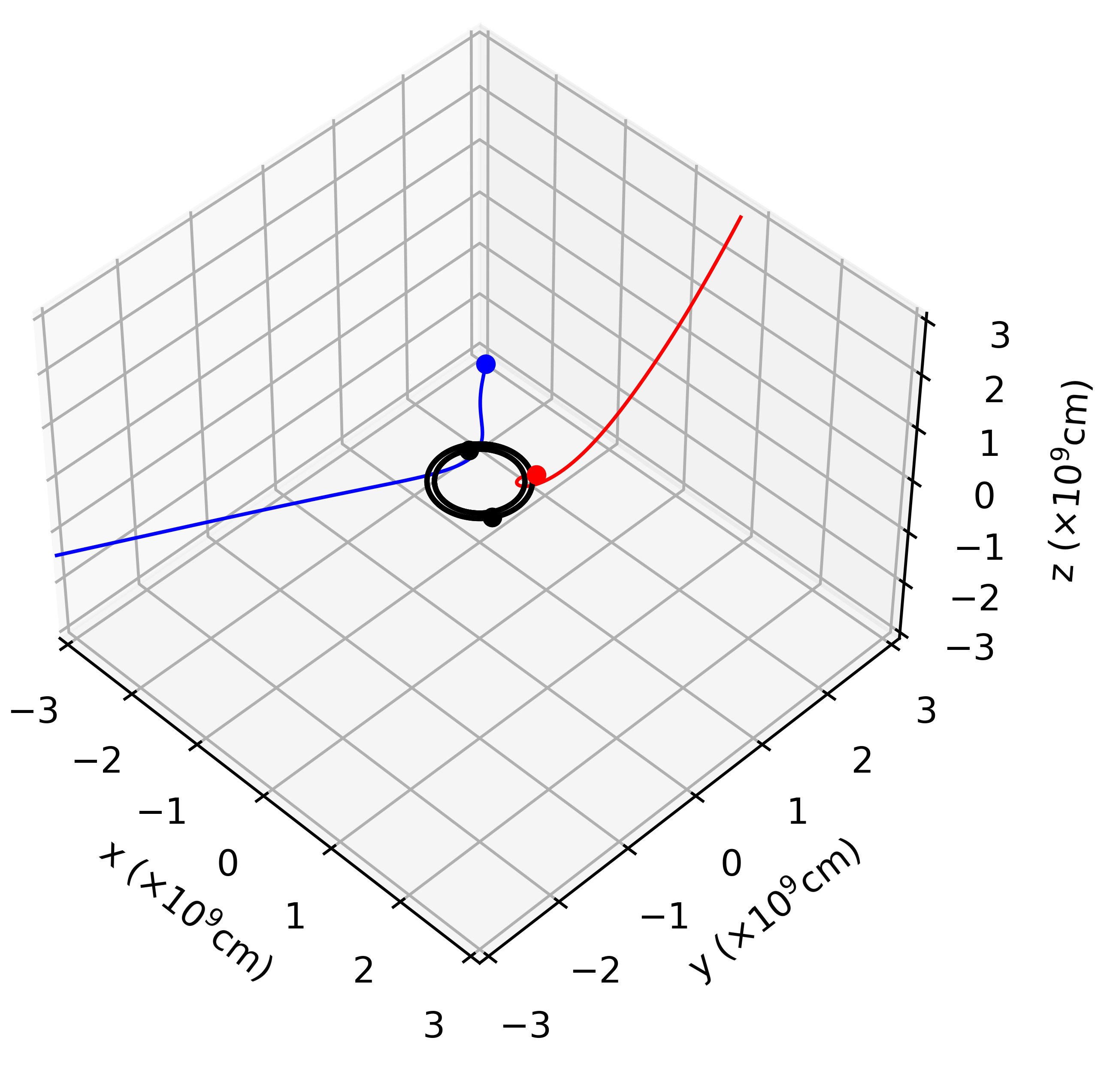}}
        \caption{\hspace{12pt}}
    \end{subfigure}
    \hfill
    \begin{subfigure}{0.45\columnwidth}
        \resizebox{0.8\hsize}{!}{\includegraphics[scale = 1.0]{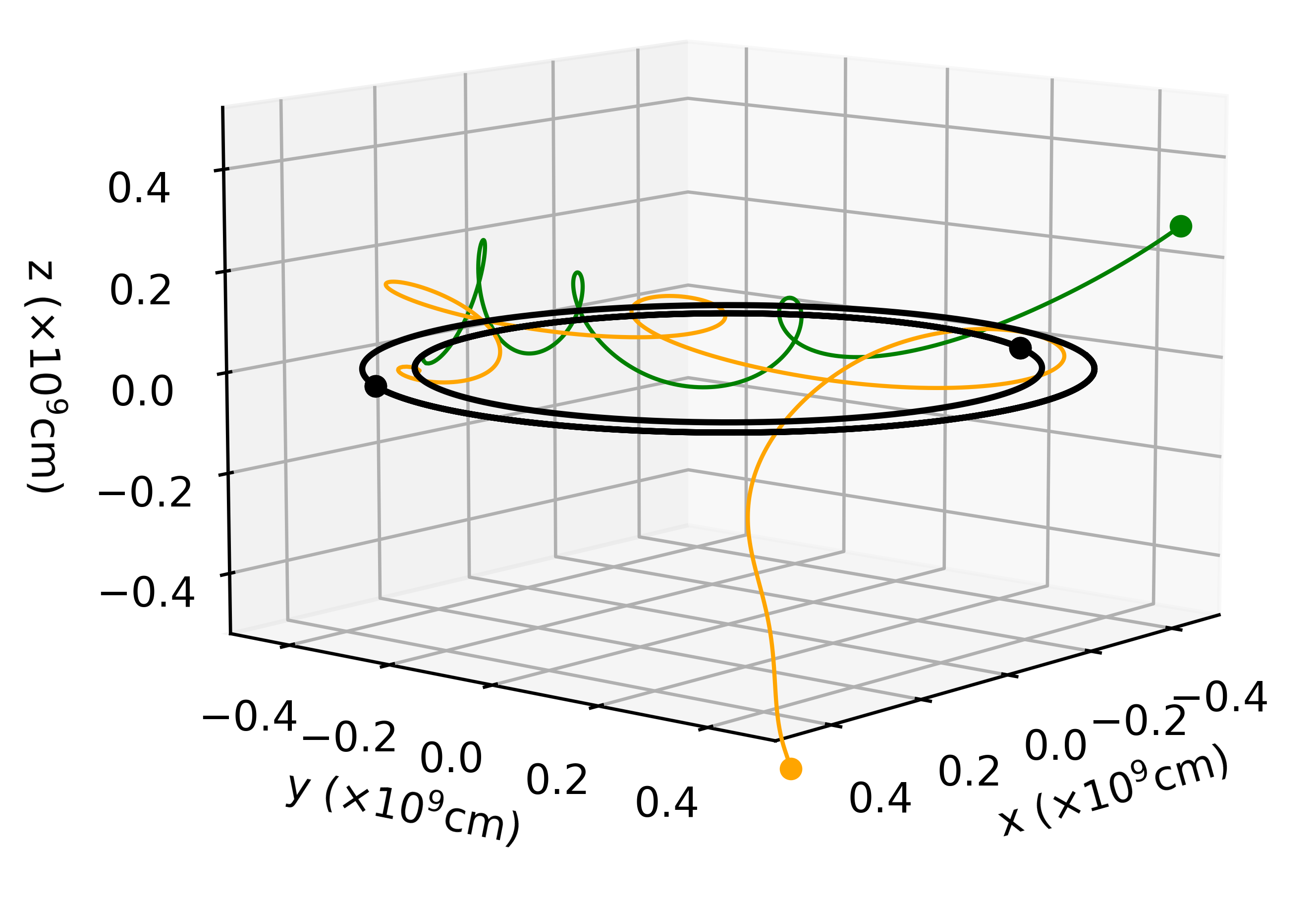}}
        \caption{\hspace{12pt}}
    \end{subfigure}
    \caption{Example of a selection of 4 particles orbiting through an inspiraling BBH system with masses $30\text{M}_{\odot}$ and $35\text{M}_{\odot}$. Two of the particles are chosen from the subset of particles that escape from the system (panel (a)), with the other two chosen from the subset of the particles that plummet into one of the BHs (panel (b)). The dots indicate the initial position of the particles. The particles used here are taken to be electrons, though the model is independent of the particle mass.}\label{fig:Orbits}%
\end{figure}
\begin{table}[!hbt]
\begin{center}
\resizebox{0.75\hsize}{!}{%
    \begin{tabular}{c c c c c}
        \hline
        Property & Particle 1 & Particle 2 & Particle 3 & Particle 4 \\ 
        \hline
        \hline
        $v_{x} (\text{cm}\cdot \text{s}^{-1})$ & $-2.52\times10^{4}$ & $2.33\times10^{3}$ & $3.80\times10^{3}$ & $-6.30\times10^{4}$ \\
        $v_{y} (\text{cm}\cdot \text{s}^{-1})$ & $-1.10\times10^{3}$ & $-1.55\times10^{3}$ & $-1.42\times10^{3}$ & $-4.33\times10^{3}$ \\
        $v_{z} (\text{cm}\cdot \text{s}^{-1})$ & $3.60\times10^{3}$ & $-4.93\times10^{3}$ & $-1.59\times10^{3}$ & $3.02\times10^{3}$ \\
        $x (\text{cm})$& $2.69\times10^{8}$ & $-8.09\times10^{8}$ & $-5.33\times10^{8}$ & $1.47\times10^{9}$ \\
        $y (\text{cm})$& $5.50\times10^{8}$ & $8.96\times10^{8}$ & $4.32\times10^{8}$ & $1.33\times10^{9}$ \\
        $z (\text{cm})$& $-1.42\times10^{8}$ & $6.64\times10^{8}$ & $2.50\times10^{8}$ & $-1.63\times10^{8}$ \\
        \hline
    \end{tabular}}   
    \caption{Initial conditions of the set of example particles. These values are randomly generated by the code for use in the model.}\label{tab:InitCon}
\end{center}
\end{table}

\begin{figure}[!hbt]
    \centering
    \begin{subfigure}{0.45\columnwidth}
        \resizebox{0.8\hsize}{!}{\includegraphics{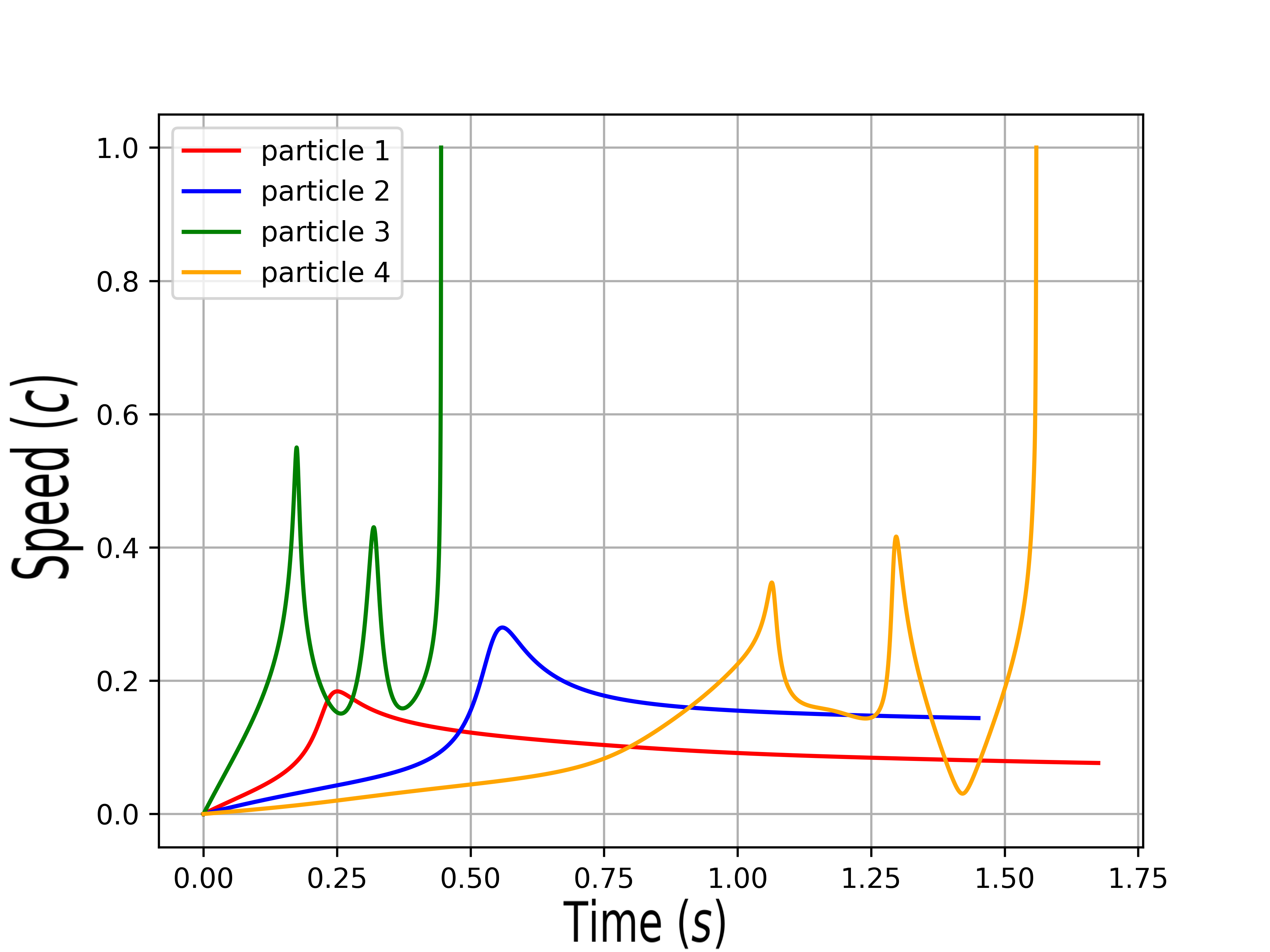}}
        \caption{\hspace{12pt}}
    \end{subfigure}
    \hfill
    \begin{subfigure}{0.45\columnwidth}
        \resizebox{0.8\hsize}{!}{\includegraphics{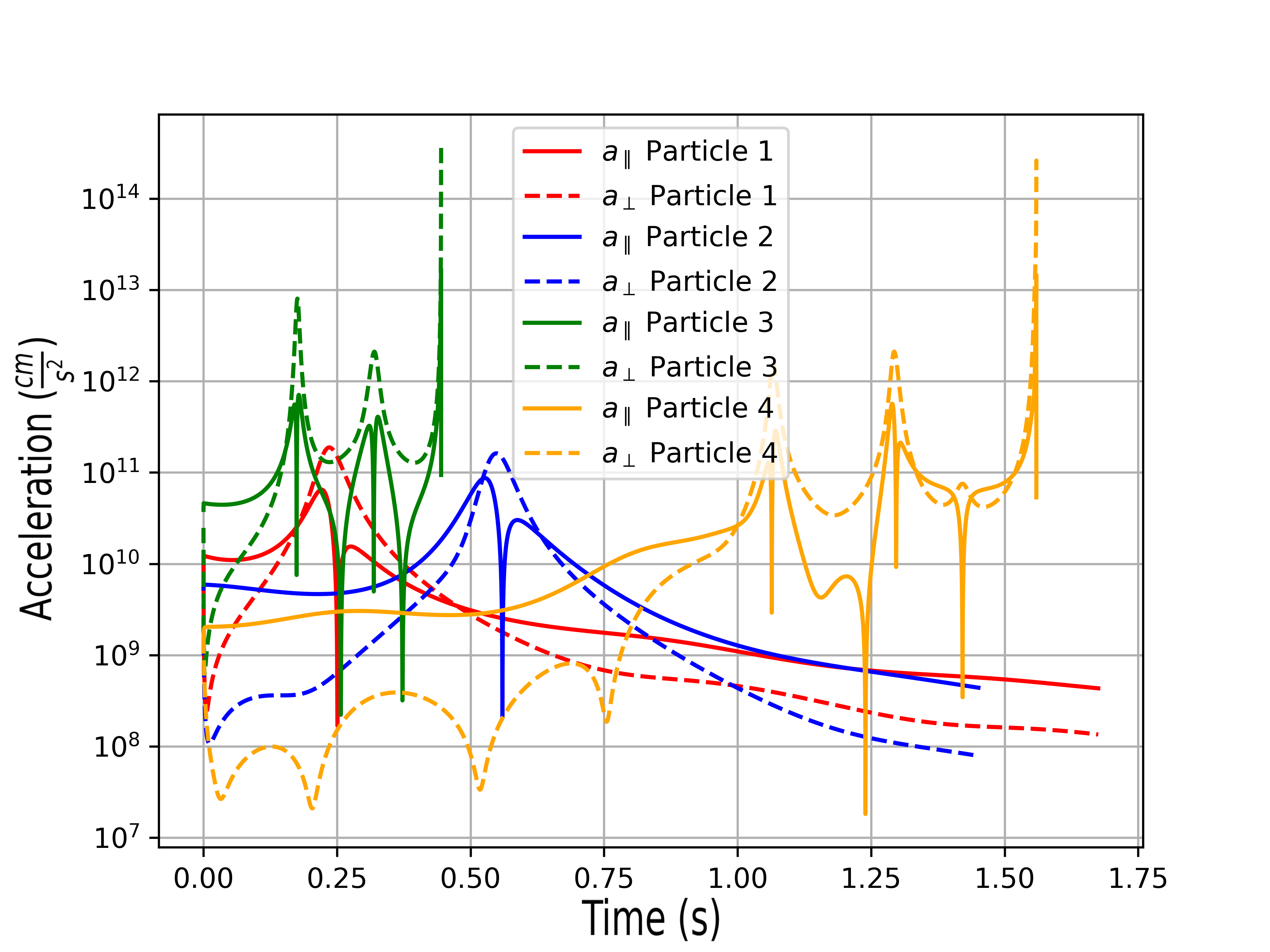}}
        \caption{\hspace{12pt}}
    \end{subfigure}
    \caption{The speed (panel (a)) and acceleration components (panel (b)) of the selection of example particles as they orbit through the BBH inpiral system. In both panels (a) and (b) the peaks correspond to the particles orbiting close to one of the BHs in the system, with the final rise of particles 3 and 4 corresponding with the particles plummeting into one of the BHs.}\label{fig:SpeedAccel}

\end{figure}

\begin{figure}[!hbt]
    \centering
    \begin{subfigure}{0.45\columnwidth}
        \resizebox{0.8\hsize}{!}{\includegraphics{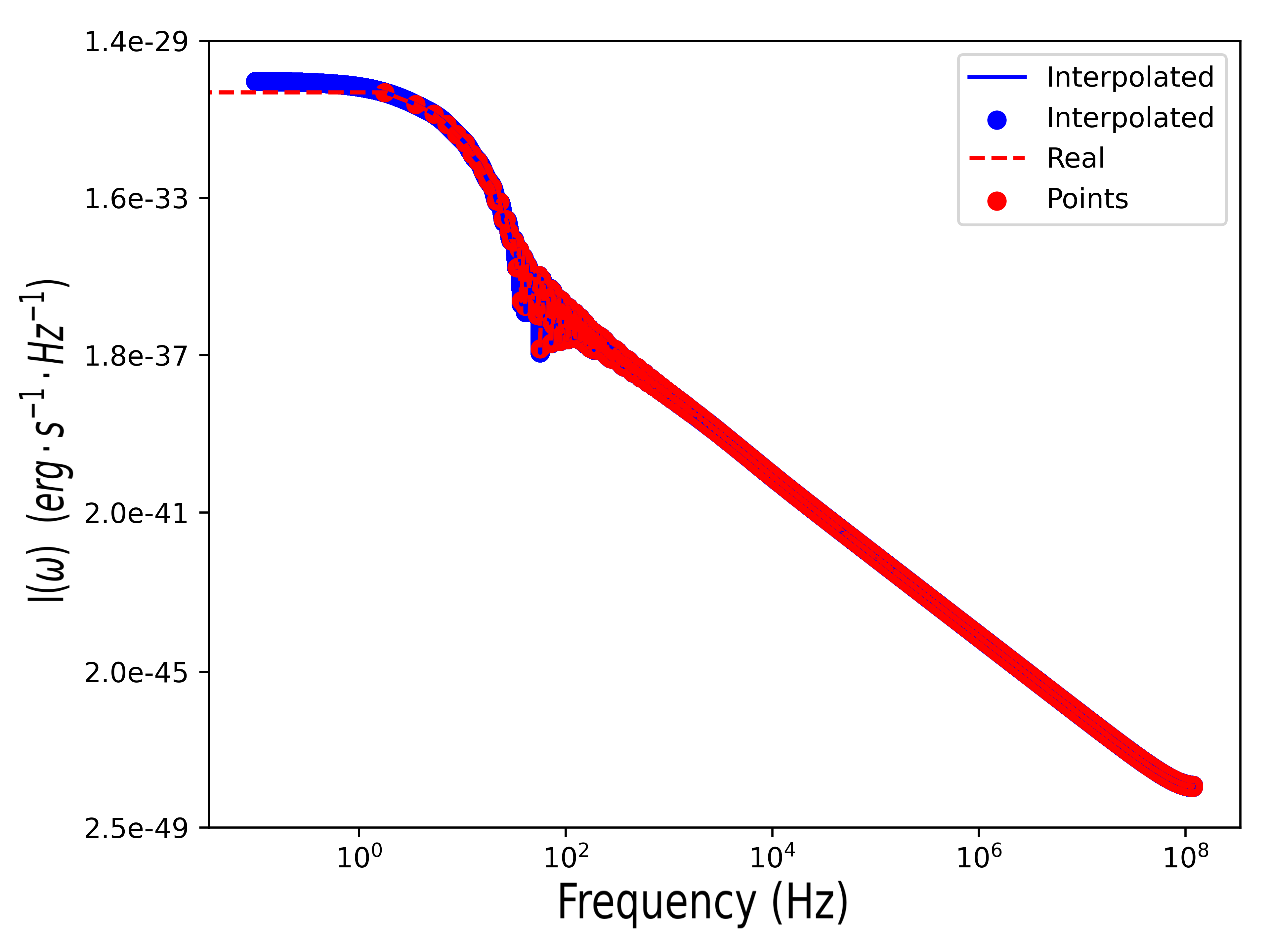}}
        \caption{\hspace{12pt}}
    \end{subfigure}
    \begin{subfigure}{0.45\columnwidth}
        \resizebox{0.8\hsize}{!}{\includegraphics{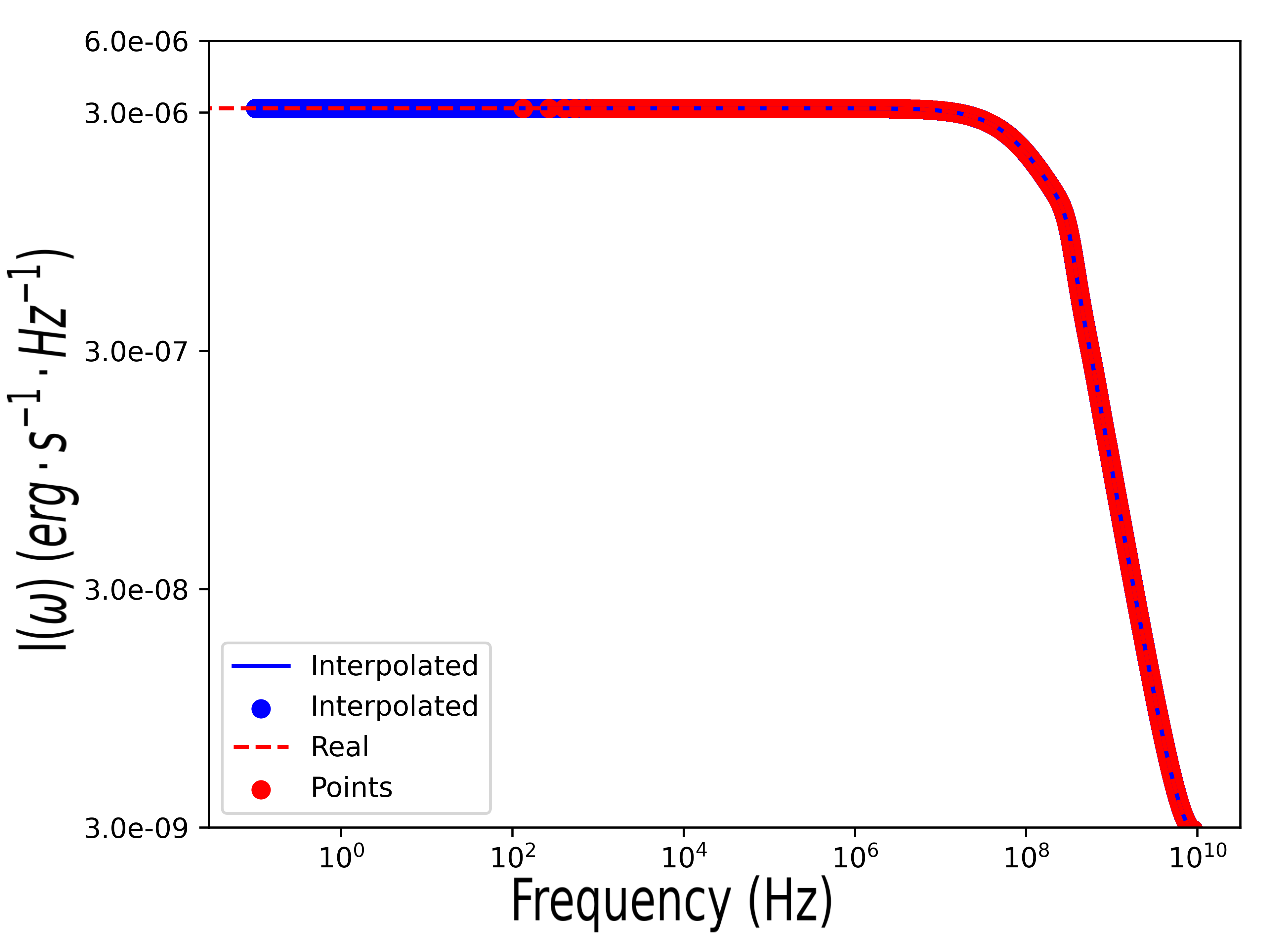}}
        \caption{\hspace{12pt}}
    \end{subfigure}
    \caption{Individual SEDs of the selection of 2 of the example particles. Panel (a) shows the SED of particle 1. This particle shows an orbit that escapes from the system and has a drop-off that occurs at around $\sim10^{1} \text{Hz}$. Panel (b) shows the SED of particle 3. This particle has an orbit that plummets into one of the BHs and shows a higher amplitude than that of particle 1. The SED of particle 3 (panel (b)) shows a drop-off at around $\sim10^{8} \text{Hz}$.}\label{fig:SEDs}
\end{figure}

\begin{figure}[!hbt]%
    \centering%
    \begin{subfigure}{0.45\columnwidth}
        \resizebox{0.8\hsize}{!}{\includegraphics[scale = 1.0]{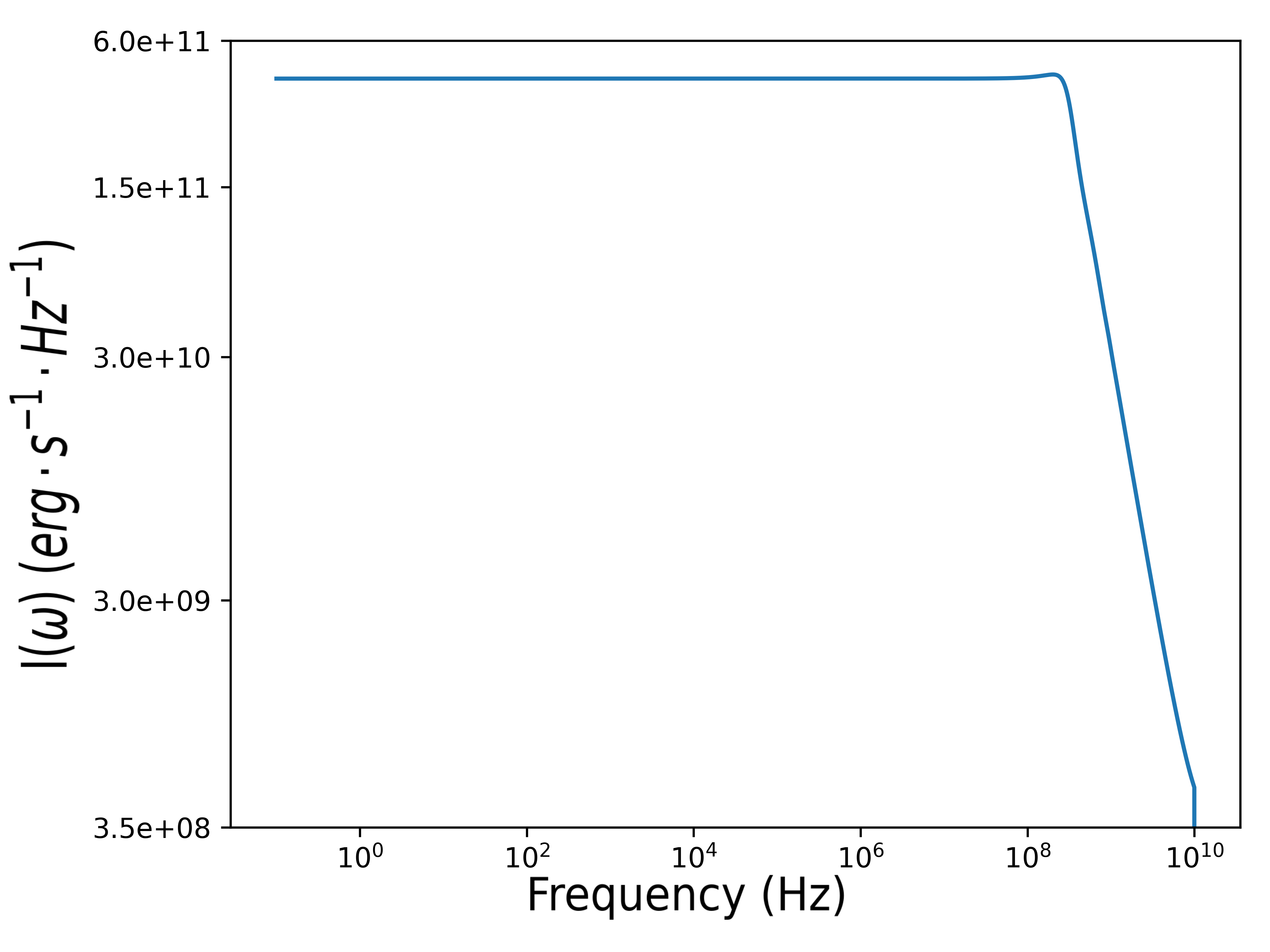}}
        \caption{Total SED from particles evolved through the inspiraling BBH system, with a steep drop-off occurring at $\sim10^{8}\text{Hz}$.\\}\label{fig:TotSED}
    \end{subfigure}
    \hfill
    \begin{subfigure}{0.45\columnwidth}
        \resizebox{0.8\hsize}{!}{\includegraphics[scale = 1.0]{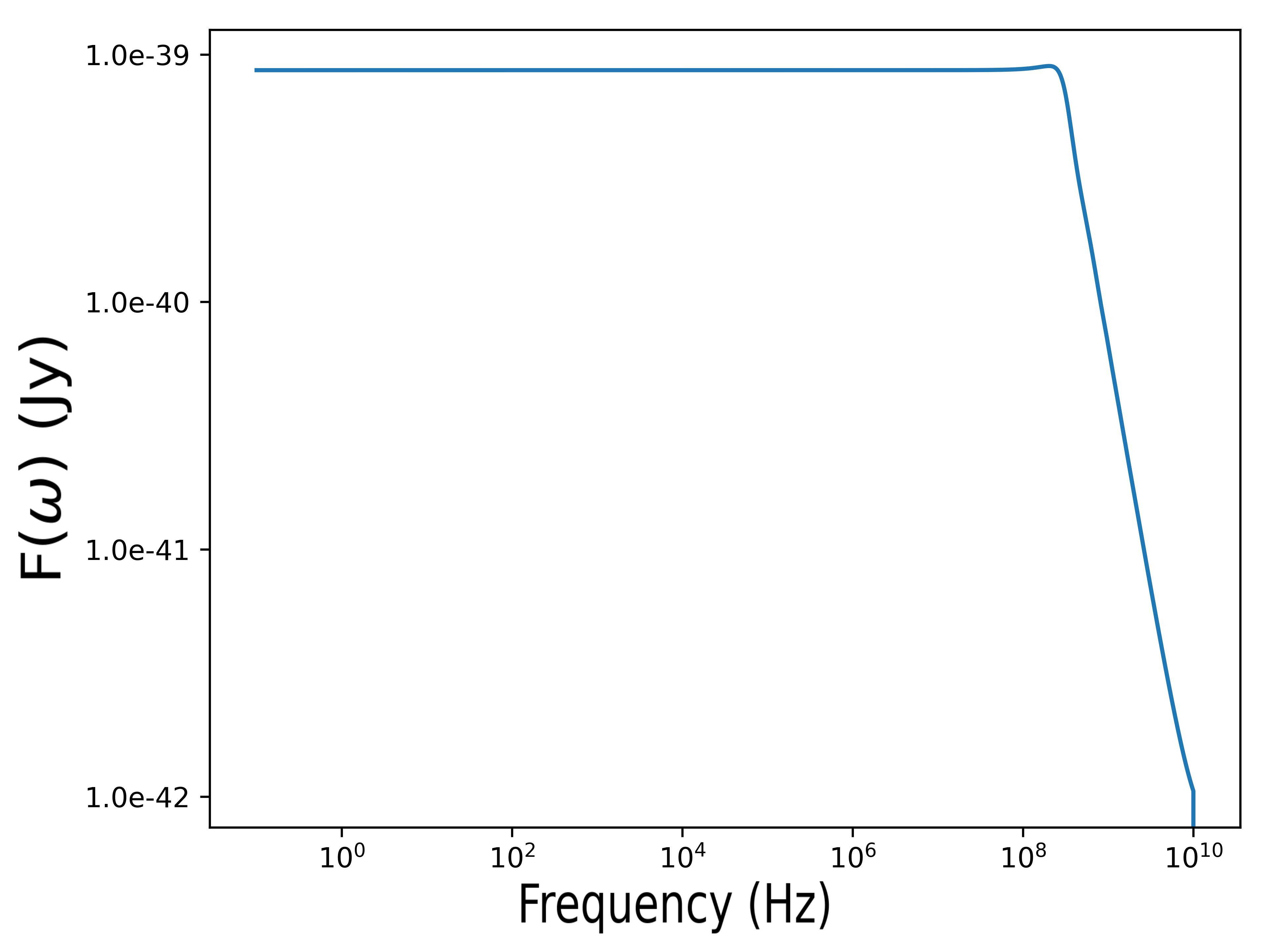}}
        \caption{Estimated flux density reaching Earth for an assumed particle density of $10^{6} \text{cm}^{-3}$ in a spherical volume of radius $R_{V} = 10^{10} \text{cm}$ for the system, with the system assumed to be a distance of $\sim400 Mpc$ from the earth.}\label{fig:EsFlux}
    \end{subfigure}
    \caption{The total SED (a) and an estimate of the flux density reaching Earth(b).}
\end{figure}

An example of a selection of the particle orbits produced by the model for BHs of masses $M_{1}=30\text{M}_{\odot}$ and $M_{2}=35\text{M}_{\odot}$, with an initial separation of $10^{9}\text{cm}$, is given in figure \ref{fig:Orbits}, with initial positions and velocities summarised in table \ref{tab:InitCon}. Of the 1060 particles evolved through the system, 150 show orbits plummeting into one of the BHs. Figure \ref{fig:SpeedAccel} shows the corresponding speed (panel (a)) and acceleration components (panel (b)) of the particles shown in figure \ref{fig:Orbits}. As is evident in panel (a) of figure \ref{fig:SpeedAccel}, the speeds of the plummeting particles approach the speed of light, while the particles that leave the system show an initial significant increase in speed, however, this stabilizes as the particles escape the system. Figure \ref{fig:SEDs} shows the individual SEDs of 2 of the selected example particles. The escaping particle SED of particle 1 (panel (a)) shows a drop-off at $\sim10^{1}\text{Hz}$ and a negligible amplitude. The SED of particle 3 (panel (b)) that plummets into one of the BHs, shows a drop-off at a much higher frequencies ($\sim10^{8}\text{Hz}$).  Figure \ref{fig:TotSED} shows the total SED of all the particles evolved through the system.\\


\section{\label{chpt:Discussion}Discussion and Conclusion}
The majority of particles are ejected from the system, leading to a significant drop-off in their radiation output at $\sim10^{1}\text{Hz}$, an example of which can be seen in figure \ref{fig:SEDs} (panel (a)). A quick order-of-magnitude estimate using the gravitational orbital frequency $\omega_{orb} = \sqrt{\frac{GM}{a^{3}}}\sim3\text{Hz}$, with $a\sim10^{9}\text{cm}$ the characteristic radius of non-plummeting orbits , reveals that this drop-off can likely be attributed to the orbital frequency of particles orbiting in the system. In contrast to this a handful of the particles plummet into one of the BHs. As can be seen in figure \ref{fig:SEDs} (panel (b)), the SEDs of these particles show a drop-off at significantly higher frequencies ($\sim10^{8}\text{Hz}$). The higher drop-off frequency for the plummeting particles can likely be attributed to the chaotic orbits of these particles on the scale of the Schwarzschild radii of the black holes. However, another order-of-magnitude estimate taking $a\sim R_{s,k}$, with $k=1,2$, shows that $\omega_{1}\sim3.3 \times 10^{3}\text{Hz}$,and $\omega_{2}\sim2.9 \times 10^{3}\text{Hz}$, indicating that the drop-off must be due to chaotic motion on even smaller scales. The total SED of the system shown in figure \ref{fig:TotSED} indicates that the SEDs of the plummeting particles completely dominate the radiative output.

If we consider that the interstellar medium (ISM) plasma frequency is $\backsim2\text{kHz}$ (for an assumed electron density of $0.03\text{cm}^{-3}$), we know that radiation emitted at frequencies below this threshold will be absorbed by the ISM. As such, only radiation emitted in the range between $\sim2\text{kHz}$ and $\sim100\text{MHz}$ will emerge from the system and be detectable from Earth. Furthermore, from figure \ref{fig:TotSED}, it is also evident that our model predicts that the bulk of possible EM radiation that originates from charged particles accelerated in a BBH merger is distributed at frequencies close to or below the lower end of the sensitivity ranges of current low-frequency radio telescopes. The current sensitivity range of the Low-Frequency Array (LOFAR) is $10-240\text{MHz}$ \citep{LOFAR2013}. As such, it may be possible to detect radiation due to charged particles accelerated in inspiraling BBH system using LOFAR or other radio telescopes operating in a similar range given a strong enough signal. In order to estimate a rough flux density $F(\omega)$ as a function of frequency for our results, we rescale the total SED in figure \ref{fig:TotSED} for $4.2\times10^{36}$ particles, based on an assumed particle density of $10^{6} \text{cm}^{-3}$ in a volume of radius $R_{V} = 10^{10} \text{cm}$ for the system\footnote{The assumed particle density assumed here is just a Highly optimistic guess based on the ISM particle density which is typically several orders of magnitude lower than this.}. Placing the system at a distance of $\sim400 \text{Mpc}$ from the earth, which is consistent with that of GW150914 \citep{GWFirst1}, the estimated flux density $F(\omega)$ can be seen in figure \ref{fig:EsFlux}. Comparing the extremely low estimated flux density of the system with the $\mu Jy$ sensitivity of LOFAR \citep{van_Haarlem_2013}, it is apparent that these events will not be detectable from Earth if only acceleration due to gravity is considered.

\FloatBarrier
\section{Acknowledgements}
This work is based on the research supported in part by the National Research Foundation of South Africa (Grant Number 144799) and in part by the National Astrophysics and Space Science Program of South Africa.



\bibliography{PoSArticle}{}
\bibliographystyle{bibliography}

\end{document}